\documentclass[12pt]{article}

\usepackage[a4paper,margin=1in]{geometry}
\usepackage{amsmath,amssymb}
\usepackage{booktabs}
\usepackage{graphicx}
\usepackage{float}
\usepackage{longtable}
\usepackage{array}
\usepackage{caption}
\usepackage{subcaption}
\usepackage{natbib}
\usepackage[colorlinks=true,linkcolor=blue,citecolor=blue,urlcolor=blue]{hyperref}

\graphicspath{{figures/}}
\newcommand{\pct}[1]{#1\%}
\newcommand{\code}[1]{\texttt{\detokenize{#1}}}

\title{\bfseries Continuous Timing Signals for Growth--Defensive Style Allocation\\
\large Factor Attribution, Risk Matching, Out-of-Sample Evidence, and a Bond/Credit Incremental Extension}
\author{Zheli Xiong\\
\normalsize Corresponding author: Zheli Xiong (zlxiong@mail.ustc.edu.cn)}
\date{}

\begin{document}
\maketitle

\begin{abstract}
This paper studies conditional allocation between a growth/technology ETF basket, denoted by $G$, and a defensive income/value-oriented ETF basket, denoted by $D$. The objective is not to discover a new standalone alpha factor, but to examine whether known style exposures can be dynamically allocated using macro-market timing signals. Fama--French five-factor plus momentum attribution shows that the relative portfolio $G-D$ is a recognizable style portfolio: its market beta is $0.273$, its HML beta is $-0.552$, its momentum beta is $0.117$, and its annualized alpha is \pct{1.95} with a Newey--West t-statistic of only $0.81$. The empirical object is therefore interpreted as a growth-versus-defensive style allocation problem rather than a new return anomaly.

The allocation framework replaces discrete regime labels and if-then trading rules with a continuous smooth score. The score combines rate relief, SPY drawdown depth, high-VIX stress relief, and a growth-crowding penalty. Interaction terms are smoothed with softplus functions, the total score is mapped to $G/D$ weights through a hyperbolic tangent function, and realized weights are smoothed with EWMA. In the main aligned comparison window from June 28, 2017 to May 15, 2026, with 10bp transaction costs, the selected smooth-score policy uses a 50\% maximum active tilt and obtains a \pct{19.24} CAGR, a Sharpe ratio of $1.01$, and a maximum drawdown of \pct{-31.63}. It improves over 50/50 $G/D$, matched TNX-only, matched core-only, SPY, and volatility-matched 100\% $G$ benchmarks. It does not, however, exceed 100\% $G$ or the best high-$G$ static portfolios in raw CAGR.

The paper further tests whether bond and credit variables add value beyond the already selected smooth-score policy. A strict incremental branch keeps the old Best Local score fixed and adds only credit relief and a rate-relief-by-credit-stress interaction. The selected incremental specification improves the old Best Local policy from a \pct{19.24} CAGR and $1.01$ Sharpe to a \pct{19.80} CAGR and $1.04$ Sharpe, while reducing annual turnover from \pct{469.67} to \pct{410.23}. Walk-forward and post-2022 validations show that the credit overlay is more useful as an incremental overlay to the original score than as a full replacement for it. Overall, the evidence supports continuous, interpretable style timing, while also showing that high static growth exposure remains a strong benchmark.
\end{abstract}


\section{Introduction}

This paper studies whether observable macro-market conditions can improve allocation between growth/technology assets and defensive income/value-oriented assets. The growth/technology basket $G$ is a fixed equal-weight portfolio of QQQ, XLK, VGT, SPYG, and VUG. The defensive income basket $D$ is a fixed equal-weight portfolio of SCHD, VYM, VTV, FDVV, and COWZ. The central relative return is
\begin{equation}
R^{G-D}_t = R^G_t - R^D_t .
\end{equation}

The research question is deliberately narrower than a search for a new alpha. Before studying conditional allocation, the paper first asks what $G-D$ is in factor terms. If $G-D$ is mostly high market beta, negative value exposure, positive momentum exposure, and aggressive investment exposure, then the relevant empirical question is not whether the portfolio is a new anomaly. It is whether these known style exposures can be managed more effectively through observable timing signals.

The empirical design has two parts. First, Fama--French five-factor plus momentum attribution identifies the systematic exposures of $G$, $D$, and $G-D$. Second, a smooth continuous score policy maps rate, drawdown, VIX, and growth-extension information into continuously varying $G/D$ weights. This structure keeps the interpretation explicit: the paper studies style timing and risk-adjusted allocation, not unexplained alpha generation.

The contribution is a transparent empirical framework for conditional style allocation. Instead of manually assigning discrete states such as ``panic,'' ``recovery,'' or ``overheated growth,'' the policy uses continuous inputs, smooth interaction functions, tanh weight mapping, and EWMA weight smoothing. This reduces the arbitrariness of threshold-based state definitions and creates a cleaner bridge between diagnostic evidence and implementable portfolio weights.

Code, data artifacts, and reproducible empirical materials are available on GitHub at:
\begin{center}
\url{https://github.com/ZheliXiong/continuous-smooth-signals-growth-tech-defensive-income-allocation}
\end{center}

\section{Related Literature}

The first literature branch is empirical asset pricing. \citet{fama1993common} and \citet{fama2015five} provide the factor-attribution framework used to distinguish systematic style exposure from residual alpha. \citet{carhart1997persistence} motivates the inclusion of momentum. In this paper, FF5 plus momentum attribution is not a side exercise; it defines the boundary of the interpretation. If $G-D$ has strong negative HML and positive momentum exposure, then the later timing exercise is more appropriately called factor or style timing.

The second branch is return predictability and out-of-sample discipline. \citet{goyal2008comprehensive} show that many in-sample return predictors fail to beat simple benchmarks out of sample. \citet{campbell2008predicting} provide a more constructive interpretation: even small out-of-sample improvements may have economic value when translated into portfolio decisions. This paper therefore separates candidate parameter discovery from validation and reports matched benchmarks, walk-forward validation, fixed-parameter validation, and a post-2022 robustness check.

The third branch is regime-based allocation and portfolio policy design. \citet{hamilton1989new} and \citet{guidolin2007asset} motivate the idea that asset behavior may vary across market states. However, rather than estimating a full latent-state model, this paper constructs transparent continuous state variables. The weight-mapping logic is also related to the parametric portfolio policy view of \citet{brandt2009parametric}, where portfolio weights are modeled as functions of observable characteristics.

Finally, the design is constrained by data-snooping concerns. \citet{white2000reality}, \citet{hansen2005superior}, and \citet{bailey2014probability} warn that repeated rule selection can overstate strategy performance. For this reason, the in-sample selected local-grid policy is interpreted as candidate discovery, while walk-forward, fixed-parameter, and post-2022 validations are used to assess the evidence more conservatively.

\section{Data and Portfolio Construction}

ETF daily returns are computed from Moomoo QFQ daily close data. For isolated missing ETF bars, prices are forward-filled for at most three trading days before computing returns; longer gaps and pre-listing periods are not filled. Factor data are Kenneth French daily FF5 factors plus daily momentum, expressed in decimal returns.

The common $G/D$ source return sample begins on December 21, 2016. Factor attribution runs from December 21, 2016 to March 31, 2026. Smooth-score policy results use source returns through May 15, 2026. The full dynamic smooth-score policy requires the 126-trading-day $G-D$ trailing return, so the aligned trading window for main comparisons begins on June 28, 2017. This is a feature warmup, not a discretionary mid-sample truncation.

\begin{table}[H]
\centering
\caption{ETF Data Coverage}
\label{tab:data-coverage}
\begin{tabular}{lllr}
\toprule
Symbol & Group & First Return Date & Number of Returns\\
\midrule
QQQ & G & 2006-05-23 & 4995\\
XLK & G & 2012-01-04 & 3580\\
VGT & G & 2012-01-04 & 3580\\
SPYG & G & 2012-01-04 & 3580\\
VUG & G & 2012-01-04 & 3580\\
SCHD & D & 2012-01-04 & 3580\\
VYM & D & 2012-01-04 & 3580\\
VTV & D & 2012-01-04 & 3580\\
FDVV & D & 2016-09-16 & 2397\\
COWZ & D & 2016-12-21 & 2330\\
\bottomrule
\end{tabular}
\end{table}

\section{Factor Attribution: What Is the $G-D$ Portfolio?}

The attribution regression is
\begin{align}
R_{i,t}-R_{f,t} = \alpha_i
&+ \beta_{MKT} MKT_t
+ \beta_{SMB} SMB_t
+ \beta_{HML} HML_t \nonumber\\
&+ \beta_{RMW} RMW_t
+ \beta_{CMA} CMA_t
+ \beta_{MOM} MOM_t
+ \varepsilon_{i,t}.
\end{align}
For $G-D$, the dependent variable is $R^G_t-R^D_t$ because the risk-free leg cancels in a zero-cost relative portfolio. Newey--West standard errors are used.

\begin{table}[H]
\centering
\caption{Full-Sample FF5+MOM Attribution, 2016-12-21 to 2026-03-31}
\label{tab:ff5-portfolios}
\scriptsize
\resizebox{\textwidth}{!}{
\begin{tabular}{lrrrrrrrrrr}
\toprule
Portfolio & $n$ & Alpha Ann. & Alpha $t_{NW}$ & MKT & SMB & HML & RMW & CMA & MOM & Adj. $R^2$\\
\midrule
G & 2330 & \pct{2.24} & 1.51 & 1.148 & -0.118 & -0.298 & 0.067 & -0.071 & 0.041 & 0.965\\
D & 2330 & \pct{0.29} & 0.24 & 0.874 & 0.019 & 0.254 & 0.088 & 0.227 & -0.076 & 0.950\\
G-D & 2330 & \pct{1.95} & 0.81 & 0.273 & -0.137 & -0.552 & -0.021 & -0.298 & 0.117 & 0.757\\
\bottomrule
\end{tabular}}
\end{table}

The results show that $G-D$ is a clear style portfolio. Its positive market beta means that long growth/technology and short defensive income carries more market risk than a neutral long-short spread might suggest. Its SMB loading is mildly negative at $-0.137$, while its strongly negative HML loading confirms that the portfolio is long growth and short value/dividend-like exposure. Its positive momentum loading means that growth/technology outperformance is partly related to momentum regimes. The alpha is economically positive but statistically weak, with a Newey--West t-statistic of only $0.81$.

ETF-level regressions confirm the internal structure of the baskets. The $G$ ETFs generally have high market beta and negative HML exposure. The $D$ ETFs generally have lower market beta, positive HML exposure, and more defensive or value-like characteristics.

\begin{table}[H]
\centering
\scriptsize
\caption{Single-ETF FF5+MOM Attribution}
\label{tab:ff5-etfs}
\resizebox{\textwidth}{!}{
\begin{tabular}{llrrrrrrrrr}
\toprule
ETF & Group & Alpha Ann. & Alpha $t_{NW}$ & MKT & SMB & HML & RMW & CMA & MOM & Adj. $R^2$\\
\midrule
QQQ & G & \pct{3.57} & 2.76 & 1.078 & -0.055 & -0.312 & 0.014 & -0.216 & 0.039 & 0.924\\
XLK & G & \pct{1.86} & 1.06 & 1.193 & -0.156 & -0.316 & 0.133 & -0.048 & 0.051 & 0.907\\
VGT & G & \pct{2.77} & 1.61 & 1.190 & -0.066 & -0.329 & 0.060 & -0.118 & 0.050 & 0.916\\
SPYG & G & \pct{0.03} & 0.04 & 1.056 & -0.137 & -0.213 & 0.086 & -0.077 & 0.037 & 0.967\\
VUG & G & \pct{0.28} & 0.37 & 1.076 & -0.114 & -0.265 & 0.029 & -0.146 & 0.006 & 0.977\\
SCHD & D & \pct{0.02} & 0.02 & 0.848 & 0.002 & 0.183 & 0.233 & 0.271 & -0.081 & 0.875\\
VYM & D & \pct{-0.88} & -0.87 & 0.862 & -0.056 & 0.259 & 0.081 & 0.238 & -0.033 & 0.927\\
VTV & D & \pct{-0.73} & -0.77 & 0.887 & -0.071 & 0.304 & 0.034 & 0.185 & -0.037 & 0.943\\
FDVV & D & \pct{-0.25} & -0.18 & 0.883 & -0.043 & 0.229 & 0.048 & 0.182 & -0.068 & 0.932\\
COWZ & D & \pct{1.48} & 0.71 & 0.934 & 0.266 & 0.223 & 0.150 & 0.224 & -0.112 & 0.880\\
\bottomrule
\end{tabular}}
\end{table}

\subsection{Rolling Attribution}

Rolling 252-day and 504-day regressions show that the exposure of $G-D$ is not constant. Table \ref{tab:rolling-periods} reports selected stage summaries. The negative HML exposure persists across regimes, while market beta and momentum exposure rise in the most recent period. These results justify the use of market-state variables in the allocation module: the object being timed is a changing style exposure rather than a fixed pure alpha.

\begin{table}[H]
\centering
\caption{$G-D$ Rolling Exposure by Market Period}
\label{tab:rolling-periods}
\scriptsize
\resizebox{\textwidth}{!}{
\begin{tabular}{lrrrrrrrr}
\toprule
Period & Alpha Ann. & MKT & SMB & HML & RMW & CMA & MOM & $R^2$\\
\midrule
COVID Rebound 2020--2021 & \pct{7.06} & 0.207 & -0.166 & -0.522 & 0.155 & -0.314 & 0.100 & 0.812\\
Rate Hike 2022 & \pct{3.77} & 0.267 & -0.252 & -0.640 & -0.013 & -0.122 & 0.113 & 0.856\\
AI Rally 2023--2024 & \pct{-2.21} & 0.275 & -0.188 & -0.542 & 0.073 & -0.486 & 0.105 & 0.809\\
Recent 2025--2026Q1 & \pct{2.61} & 0.398 & -0.103 & -0.682 & -0.022 & -0.021 & 0.306 & 0.794\\
\bottomrule
\end{tabular}}
\end{table}

\section{Smooth Continuous Score Policy}

The smooth-score policy is designed to avoid hard state definitions. All variables are direction-normalized:
\begin{align}
r_t &= -z(\Delta TNX_{21,t}),\\
d_t &= -z(SPYDrawdown_t),\\
vh_t &= z(VIXPercentile_{756,t}),\\
vr_t &= -z(\Delta VIX_{21,t}),\\
g126_t &= z(GDTrailing126_t).
\end{align}
Thus, larger $r_t$ means more rate relief, larger $d_t$ means deeper drawdown, larger $vh_t$ means higher VIX, larger $vr_t$ means stronger VIX relief, and larger $g126_t$ means stronger medium-term growth outperformance.

Softplus is used to avoid hard positive-part rules:
\begin{equation}
softplus_{\tau}(x)=\tau\log(1+\exp(x/\tau)),
\end{equation}
with $\tau=1.0$ in the empirical design. The smooth components are
\begin{align}
HighVIX_t &= softplus(vh_t),\\
VIXRelief_t &= softplus(vr_t),\\
LowVIX_t &= softplus(-vh_t),\\
GrowthExt_t &= softplus(g126_t),\\
RateQuiet_t &= \exp(-0.5r_t^2).
\end{align}

The four interaction terms are
\begin{align}
i1_t &= r_t \cdot vh_t,\\
i2_t &= HighVIX_t \cdot VIXRelief_t,\\
i3_t &= GrowthExt_t \cdot LowVIX_t,\\
i4_t &= GrowthExt_t \cdot LowVIX_t \cdot RateQuiet_t.
\end{align}
The first two terms capture stress relief. The last two terms penalize crowded growth exposure when growth has already run and VIX is low, especially when rates are quiet rather than supportive.

The policy score is
\begin{align}
CoreScore_t &= \alpha r_t + (1-\alpha)d_t,\\
StressScore_t &= 0.5z(i1_t)+0.5z(i2_t),\\
CrowdedScore_t &= 0.5z(i3_t)+0.5z(i4_t),\\
RawScore_t &= CoreScore_t + \lambda_s StressScore_t - \lambda_c CrowdedScore_t.
\end{align}
The raw score is standardized again with an expanding z-score, denoted by $\widetilde{Score}_t$. The target $G$ weight is
\begin{equation}
w^{target}_{G,t}=0.5+MaxTilt\cdot \tanh(\widetilde{Score}_t/\tau_w).
\end{equation}
Actual weights are smoothed:
\begin{equation}
w_{G,t}=(1-\eta)w_{G,t-1}+\eta w^{target}_{G,t}.
\end{equation}

Signals are formed after the close on day $t$ and applied from day $t+1$. The main transaction cost is 10bp, applied as
\begin{equation}
Cost_t=2|\Delta w_{G,t}|\cdot \frac{cost\_bps}{10000}.
\end{equation}

\section{Empirical Evidence on Smooth-Score Allocation}

\subsection{Active-Tilt Sensitivity}

The first experiment fixes $\alpha=0.50$, $\lambda_s=0.25$, $\lambda_c=0.15$, $\tau_w=1.0$, and $\eta=0.05$, and varies $MaxTilt\in\{20\%,30\%,40\%,50\%\}$. Table \ref{tab:tilt} reports the 10bp results.

\begin{table}[H]
\centering
\caption{Fixed-Structure MaxTilt Test, 10bp Cost}
\label{tab:tilt}
\begin{tabular}{rrrrrrrr}
\toprule
MaxTilt & Final Wealth & CAGR & Sharpe & Max DD & Calmar & Turnover & Avg $G$\\
\midrule
\pct{20} & 4.30 & \pct{17.89} & 0.95 & \pct{-32.84} & 0.54 & \pct{186.35} & \pct{48.77}\\
\pct{30} & 4.42 & \pct{18.27} & 0.97 & \pct{-32.47} & 0.56 & \pct{279.52} & \pct{48.15}\\
\pct{40} & 4.55 & \pct{18.65} & 0.98 & \pct{-32.10} & 0.58 & \pct{372.70} & \pct{47.54}\\
\pct{50} & 4.68 & \pct{19.02} & 0.99 & \pct{-31.72} & 0.60 & \pct{465.87} & \pct{46.92}\\
\bottomrule
\end{tabular}
\end{table}

The result shows that a 20\% maximum tilt is too conservative for this score structure. Larger active tilts allow the signal to express stronger conditional views. The 50\% tilt version has higher CAGR, higher Sharpe, and lower maximum drawdown, although turnover rises materially. The subsequent comparisons therefore use the 50\% tilt family as the main candidate rather than the conservative 20\% baseline.

\begin{figure}[H]
\centering
\includegraphics[width=0.95\linewidth]{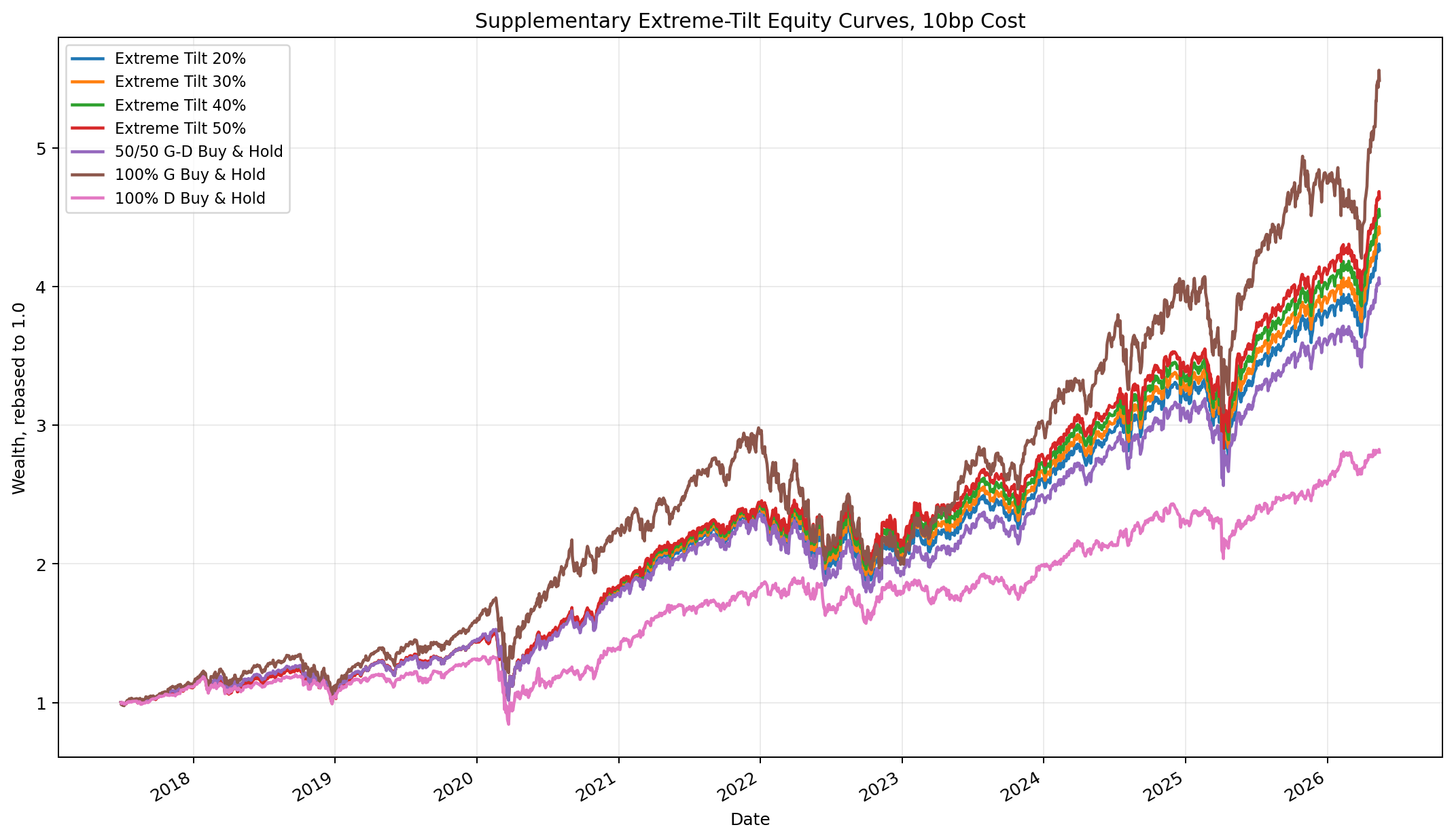}
\caption{Active-Tilt Sensitivity Equity Curves, 2017-06-28 to 2026-05-15}
\label{fig:tilt-curves}
\end{figure}

\subsection{Candidate Smooth-Score Policy Selection}

The expanded local grid tests
\begin{align}
\alpha &\in \{0.50,0.67\},\\
\lambda_s &\in \{0.25,0.50\},\\
\lambda_c &\in \{0.05,0.15,0.25\},\\
MaxTilt &\in \{20\%,30\%,40\%,50\%\},\\
\tau_w &\in \{0.75,1.0,1.5\},\\
\eta &\in \{0.03,0.05,0.10\}.
\end{align}
The selected configuration is
\begin{equation}
\alpha=0.50,\quad \lambda_s=0.50,\quad \lambda_c=0.05,\quad MaxTilt=50\%,\quad \tau_w=0.75,\quad \eta=0.05.
\end{equation}

\begin{table}[H]
\centering
\scriptsize
\caption{Expanded Local Grid, Top Five Configurations at 10bp}
\label{tab:local-grid}
\begin{tabular}{lrrrrrr}
\toprule
Config & Final Wealth & CAGR & Sharpe & Max DD & Turnover & Avg $G$\\
\midrule
\code{a0.50 ls0.50 lc0.05 tilt0.50 tau0.75 eta0.05} & 4.76 & \pct{19.24} & 1.01 & \pct{-31.63} & \pct{469.67} & \pct{45.06}\\
\code{a0.50 ls0.50 lc0.05 tilt0.50 tau1.00 eta0.05} & 4.70 & \pct{19.08} & 1.00 & \pct{-31.69} & \pct{413.13} & \pct{46.47}\\
\code{a0.50 ls0.25 lc0.05 tilt0.50 tau0.75 eta0.05} & 4.78 & \pct{19.30} & 1.01 & \pct{-31.54} & \pct{520.00} & \pct{46.99}\\
\code{a0.50 ls0.25 lc0.05 tilt0.50 tau1.00 eta0.05} & 4.72 & \pct{19.13} & 1.00 & \pct{-31.61} & \pct{459.90} & \pct{48.00}\\
\code{a0.50 ls0.50 lc0.15 tilt0.50 tau0.75 eta0.05} & 4.72 & \pct{19.14} & 1.00 & \pct{-31.75} & \pct{481.06} & \pct{44.05}\\
\bottomrule
\end{tabular}
\end{table}

The selected configuration is not simply the highest raw CAGR in the local grid. It is chosen by a composite selection score balancing Sharpe, Calmar, CAGR, drawdown, and turnover. This matters because larger active tilts can mechanically raise exposure to growth risk, while faster weight updates can increase transaction costs. At 20bp transaction cost, lower-$\eta$ versions rank higher because turnover becomes more important. The selected 10bp configuration remains above the 50/50 benchmark even under 20bp stress cost.

\begin{figure}[H]
\centering
\includegraphics[width=0.95\linewidth]{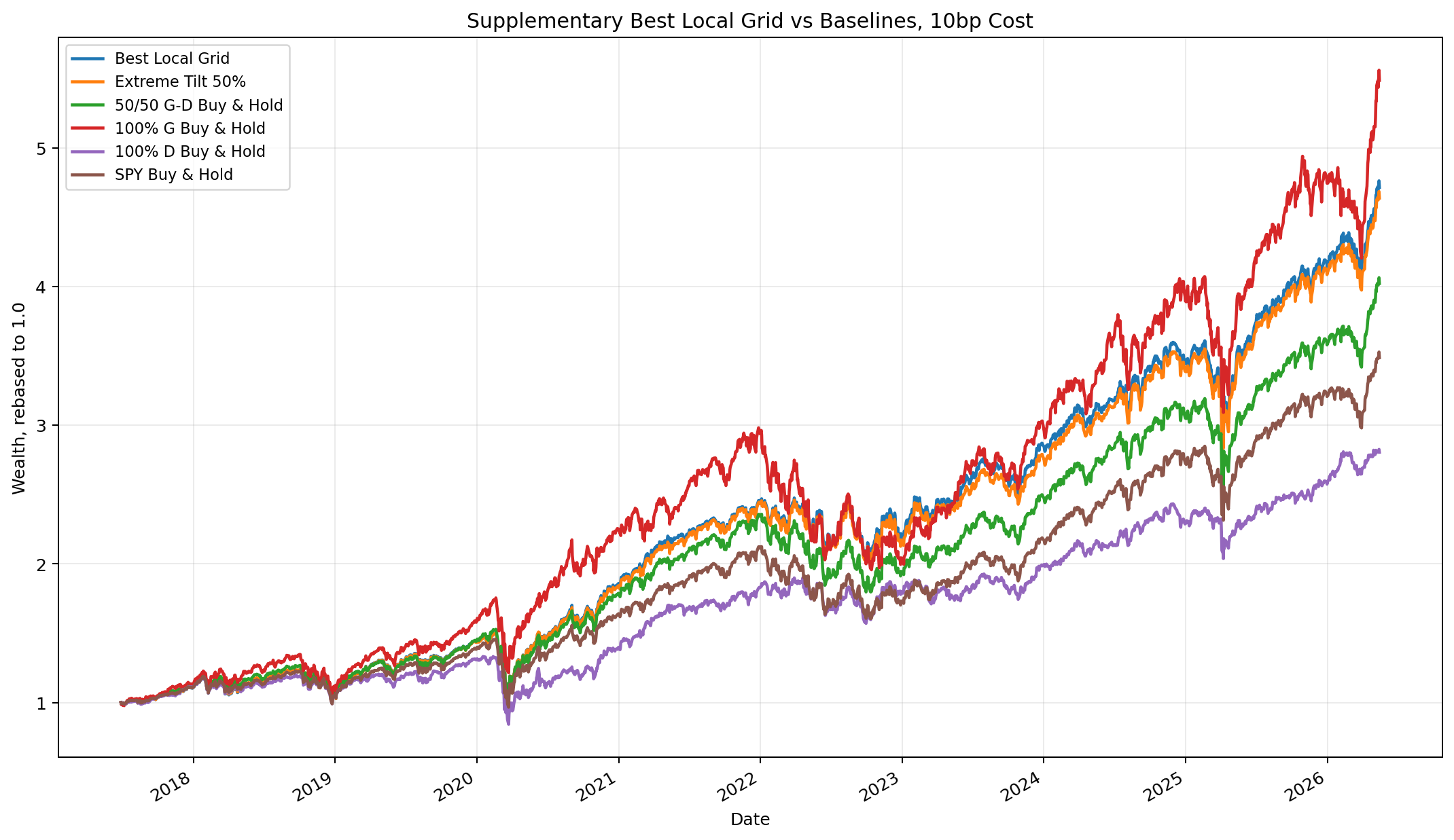}
\caption{Selected Smooth-Score Policy Equity Curves, 2017-06-28 to 2026-05-15}
\label{fig:local-curves}
\end{figure}

\section{Benchmark Evidence}

All methods in Table \ref{tab:selected-summary} are aligned from June 28, 2017 to May 15, 2026. The selected smooth-score policy is compared with matched TNX-only and matched core-only policies. These matched policies use the same $MaxTilt$, $\tau_w$, and $\eta$, but use only rate relief or rate relief plus drawdown depth. This isolates the incremental contribution of the stress-relief and crowding interactions.

\begin{table}[H]
\centering
\scriptsize
\caption{Aligned Strategy Comparison, 2017-06-28 to 2026-05-15, 10bp Cost}
\label{tab:selected-summary}
\begin{tabular}{lrrrrrrrr}
\toprule
Method & Final Wealth & CAGR & Vol & Sharpe & Sortino & Max DD & Turnover & Avg $G$\\
\midrule
Selected Smooth Score & 4.71 & \pct{19.24} & \pct{19.29} & 1.01 & 1.22 & \pct{-31.63} & \pct{469.67} & \pct{45.06}\\
Matched TNX-only & 4.23 & \pct{17.80} & \pct{19.33} & 0.94 & 1.13 & \pct{-31.31} & \pct{566.69} & \pct{47.48}\\
Matched Core-only & 4.25 & \pct{17.84} & \pct{18.94} & 0.96 & 1.15 & \pct{-31.98} & \pct{410.19} & \pct{37.18}\\
Fixed-Structure 50\% Tilt & 4.64 & \pct{19.02} & \pct{19.45} & 0.99 & 1.20 & \pct{-31.72} & \pct{465.87} & \pct{46.92}\\
50/50 $G/D$ & 4.02 & \pct{17.12} & \pct{19.34} & 0.91 & 1.08 & \pct{-33.59} & \pct{0.00} & \pct{50.00}\\
100\% $G$ & 5.49 & \pct{21.34} & \pct{23.53} & 0.94 & 1.17 & \pct{-34.35} & \pct{0.00} & \pct{100.00}\\
100\% $D$ & 2.80 & \pct{12.42} & \pct{17.53} & 0.76 & 0.86 & \pct{-36.71} & \pct{0.00} & \pct{0.00}\\
SPY & 3.49 & \pct{15.25} & \pct{18.74} & 0.85 & 0.98 & \pct{-33.72} & \pct{0.00} & --\\
\bottomrule
\end{tabular}
\end{table}

The selected policy improves over 50/50 $G/D$, matched TNX-only, matched core-only, and SPY. It does not exceed 100\% $G$ in raw CAGR. The correct interpretation is therefore not that the smooth score replaces high growth exposure, but that it improves risk-adjusted allocation and drawdown control relative to more comparable allocation benchmarks.

\begin{table}[H]
\centering
\caption{Incremental Comparison Against Key Benchmarks}
\label{tab:incremental}
\begin{tabular}{lrrrr}
\toprule
Comparison & Annual Excess & Tracking Error & Info Ratio & Max DD Diff\\
\midrule
Selected Policy -- Matched TNX-only & \pct{1.21} & \pct{2.49} & 0.48 & \pct{-0.33}\\
Selected Policy -- Matched Core-only & \pct{1.25} & \pct{1.81} & 0.69 & \pct{0.35}\\
Selected Policy -- Fixed-Structure 50\% Tilt & \pct{0.15} & \pct{1.00} & 0.15 & \pct{0.09}\\
Selected Policy -- 50/50 & \pct{1.78} & \pct{3.74} & 0.48 & \pct{1.95}\\
Selected Policy -- 100\% $G$ & \pct{-2.66} & \pct{9.48} & -0.28 & \pct{2.71}\\
Selected Policy -- SPY & \pct{3.51} & \pct{4.32} & 0.81 & \pct{2.08}\\
\bottomrule
\end{tabular}
\end{table}

\begin{figure}[H]
\centering
\includegraphics[width=0.95\linewidth]{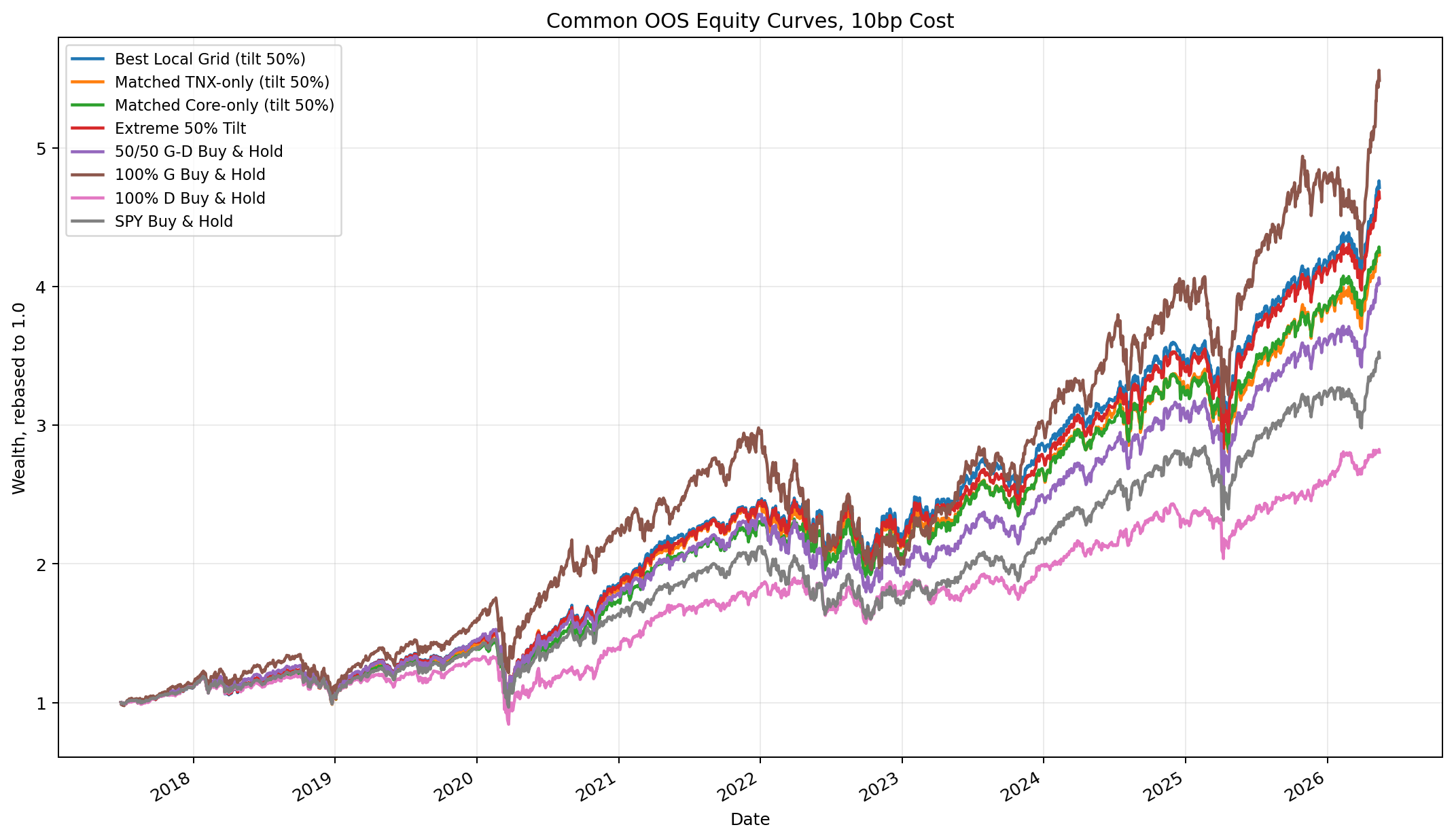}
\caption{Aligned Main Equity Curves, 2017-06-28 to 2026-05-15}
\label{fig:main-curves}
\end{figure}

\begin{figure}[H]
\centering
\includegraphics[width=0.95\linewidth]{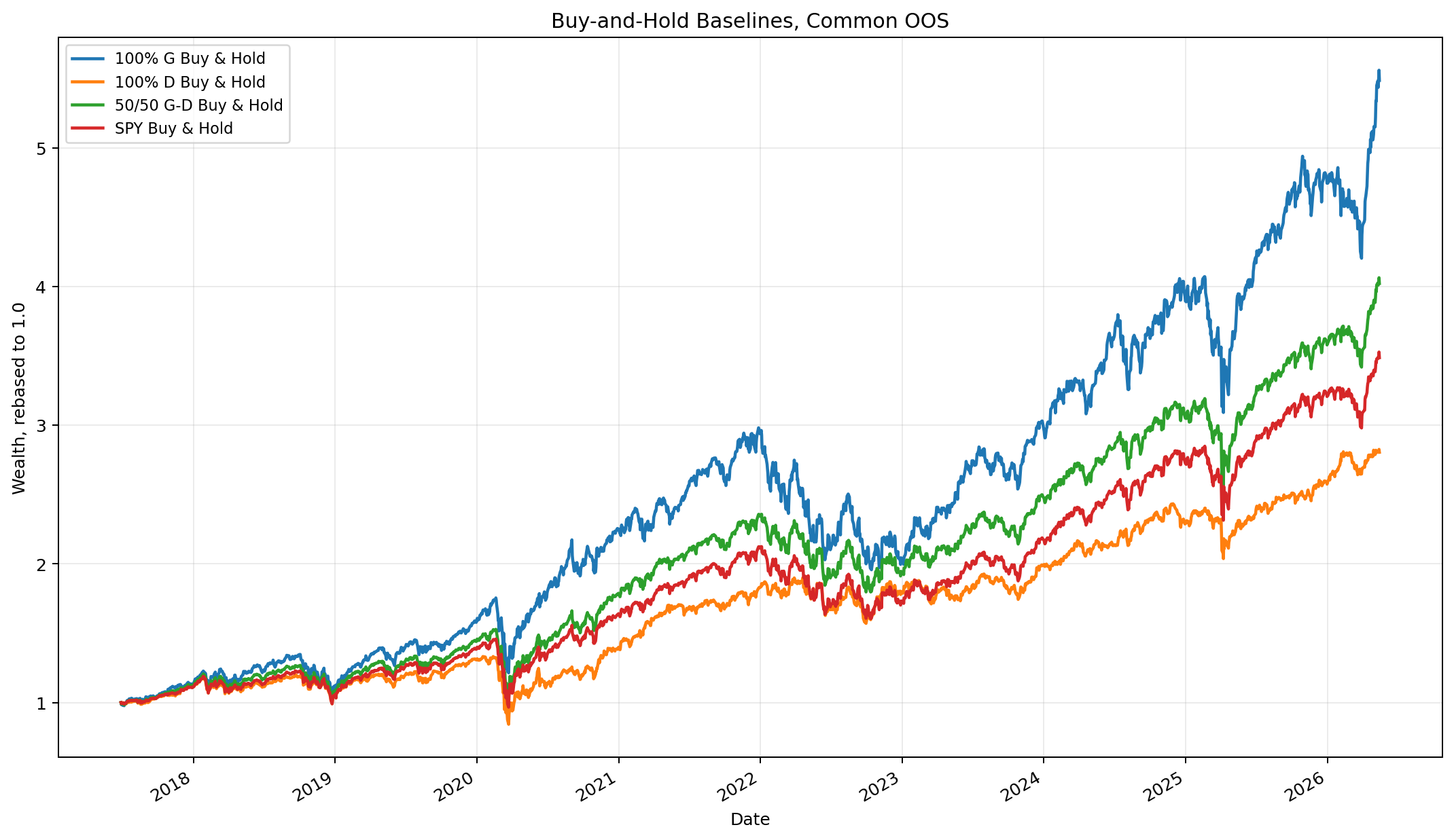}
\caption{Buy-and-Hold $G/D$ Baseline Equity Curves, 2017-06-28 to 2026-05-15}
\label{fig:buy-hold}
\end{figure}

\section{Volatility-Matched and Static $G/D$ Comparisons}

The selected smooth policy has much lower average $G$ exposure than 100\% $G$. A direct comparison of CAGR therefore mixes signal quality with risk level. The vol-matched comparison scales 100\% $G$ to the smooth strategy's annualized volatility. The required scaling weight is 81.95\%.

\begin{table}[H]
\centering
\scriptsize
\caption{Vol-Matched and Static $G/D$ Comparisons}
\label{tab:vol-matched}
\begin{tabular}{lrrrrrr}
\toprule
Method & CAGR & Vol & Sharpe & Max DD & Turnover & Excess vs Smooth\\
\midrule
Selected Smooth Score & \pct{19.24} & \pct{19.29} & 1.01 & \pct{-31.63} & \pct{469.67} & \pct{0.00}\\
100\% $G$ & \pct{21.34} & \pct{23.53} & 0.94 & \pct{-34.35} & \pct{0.00} & \pct{2.66}\\
Vol-Matched 100\% $G$ & \pct{17.66} & \pct{19.29} & 0.94 & \pct{-28.73} & \pct{0.00} & \pct{-1.34}\\
50/50 $G/D$ & \pct{17.12} & \pct{19.34} & 0.91 & \pct{-33.59} & \pct{0.00} & \pct{-1.78}\\
Vol-Matched Static $G/D$ (49\% $G$) & \pct{17.03} & \pct{19.28} & 0.91 & \pct{-33.64} & \pct{0.00} & \pct{-1.87}\\
MaxDD-Matched Static $G/D$ (87\% $G$) & \pct{20.29} & \pct{22.28} & 0.94 & \pct{-31.65} & \pct{0.00} & \pct{1.50}\\
Best Sharpe Static $G/D$ (89\% $G$) & \pct{20.46} & \pct{22.46} & 0.94 & \pct{-32.07} & \pct{0.00} & \pct{1.68}\\
\bottomrule
\end{tabular}
\end{table}

The smooth score exceeds vol-matched 100\% $G$ at the same annualized volatility. It also exceeds the 50/50 and vol-matched static $G/D$ portfolios. However, high-growth static portfolios around 87--89\% $G$ remain strong competitors in raw CAGR. This reinforces the main interpretation: the policy improves risk-adjusted allocation efficiency, but it does not eliminate the challenge posed by high static growth exposure in a strong growth sample.

\begin{figure}[H]
\centering
\includegraphics[width=0.95\linewidth]{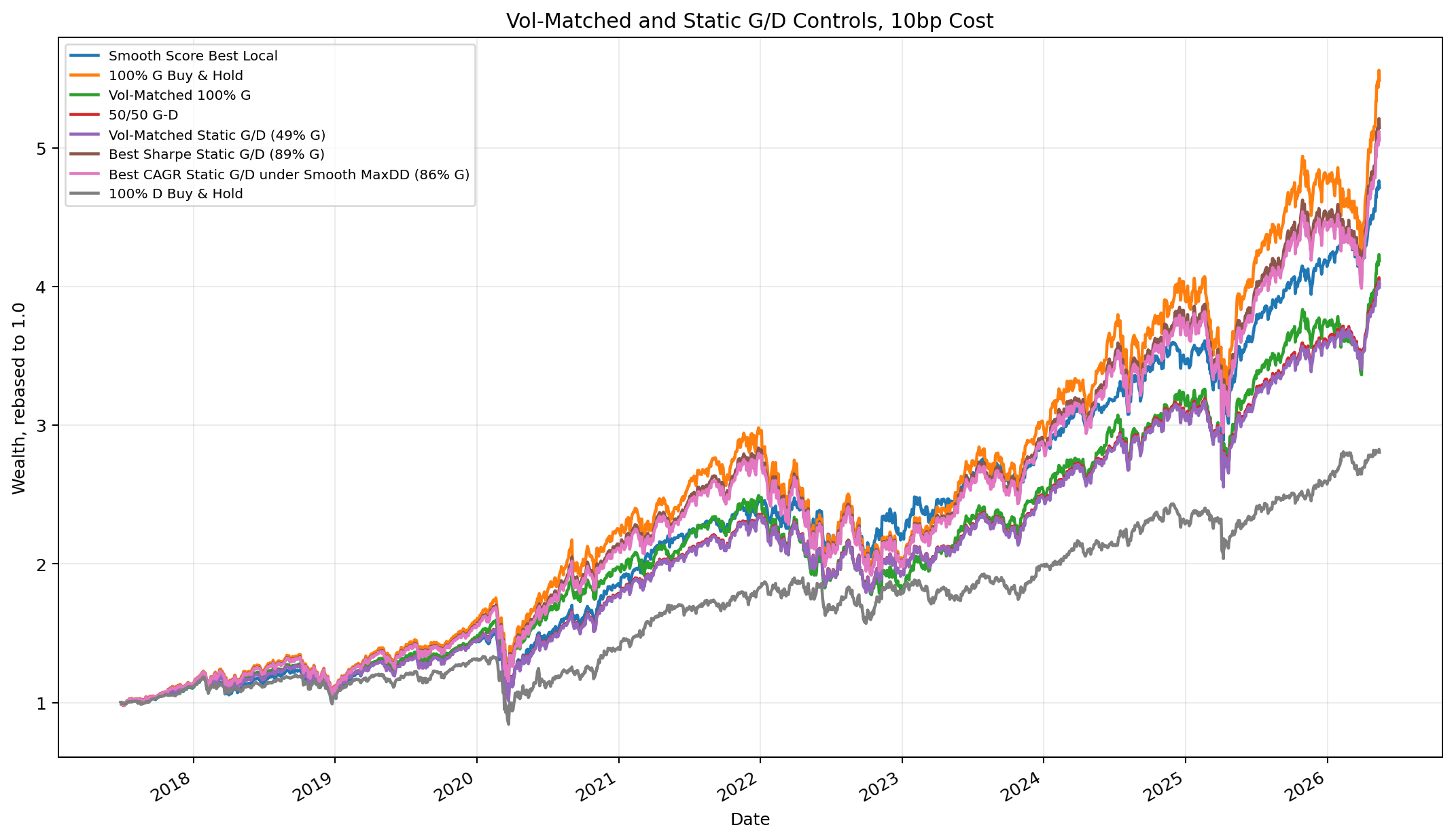}
\caption{Vol-Matched and Static $G/D$ Equity Curves, 2017-06-28 to 2026-05-15}
\label{fig:vol-matched}
\end{figure}

\section{Out-of-Sample Validation}

The validation section combines walk-forward expanding, walk-forward rolling, and fixed-parameter validation into one aligned OOS window. The parameter pool is the candidate local grid described above. The initial training window is 252 trading days, and the test block length is 63 trading days.

Walk-forward expanding uses all history available before each test block. Walk-forward rolling uses only the most recent 252 trading days before each test block. Fixed parameter selects once from the first training window, June 28, 2017 to June 27, 2018, and then trades that fixed configuration from June 28, 2018 onward. The fixed selected configuration is
\begin{equation}
\alpha=0.50,\quad \lambda_s=0.50,\quad \lambda_c=0.25,\quad MaxTilt=20\%,\quad \tau_w=1.50,\quad \eta=0.03.
\end{equation}

\begin{table}[H]
\centering
\scriptsize
\caption{OOS Validation, 2018-06-28 to 2026-05-15, 10bp Cost}
\label{tab:oos}
\begin{tabular}{lrrrrrrrr}
\toprule
Method & Final Wealth & CAGR & Vol & Sharpe & Sortino & Max DD & Turnover & Avg $G$\\
\midrule
Smooth Score WF Expanding & 3.83 & \pct{18.64} & \pct{19.77} & 0.96 & 1.16 & \pct{-32.93} & \pct{332.21} & \pct{44.82}\\
Smooth Score WF Rolling & 3.68 & \pct{18.02} & \pct{19.86} & 0.93 & 1.12 & \pct{-32.79} & \pct{350.79} & \pct{46.66}\\
Fixed Parameter & 3.64 & \pct{17.86} & \pct{19.90} & 0.93 & 1.11 & \pct{-33.19} & \pct{88.36} & \pct{48.63}\\
50/50 $G/D$ & 3.46 & \pct{17.10} & \pct{20.00} & 0.89 & 1.06 & \pct{-33.59} & \pct{0.00} & \pct{50.00}\\
100\% $G$ & 4.51 & \pct{21.13} & \pct{24.38} & 0.91 & 1.13 & \pct{-34.35} & \pct{0.00} & \pct{100.00}\\
100\% $D$ & 2.53 & \pct{12.51} & \pct{18.14} & 0.74 & 0.85 & \pct{-36.71} & \pct{0.00} & \pct{0.00}\\
SPY & 3.09 & \pct{15.45} & \pct{19.39} & 0.84 & 0.98 & \pct{-33.72} & \pct{0.00} & --\\
\bottomrule
\end{tabular}
\end{table}

The expanding walk-forward policy is the strongest OOS smooth-score variant. Rolling validation is weaker, likely because the 252-day rolling window is more sensitive to noisy recent regimes. Fixed parameter has much lower turnover but also lower performance. All three smooth-score OOS variants improve over 50/50 and SPY in this aligned OOS window. 100\% $G$ still has the highest CAGR, but with higher volatility and deeper drawdown. This comparison is consistent with a risk-adjusted style-allocation interpretation rather than a claim of unconditional dominance.

\begin{figure}[H]
\centering
\includegraphics[width=0.95\linewidth]{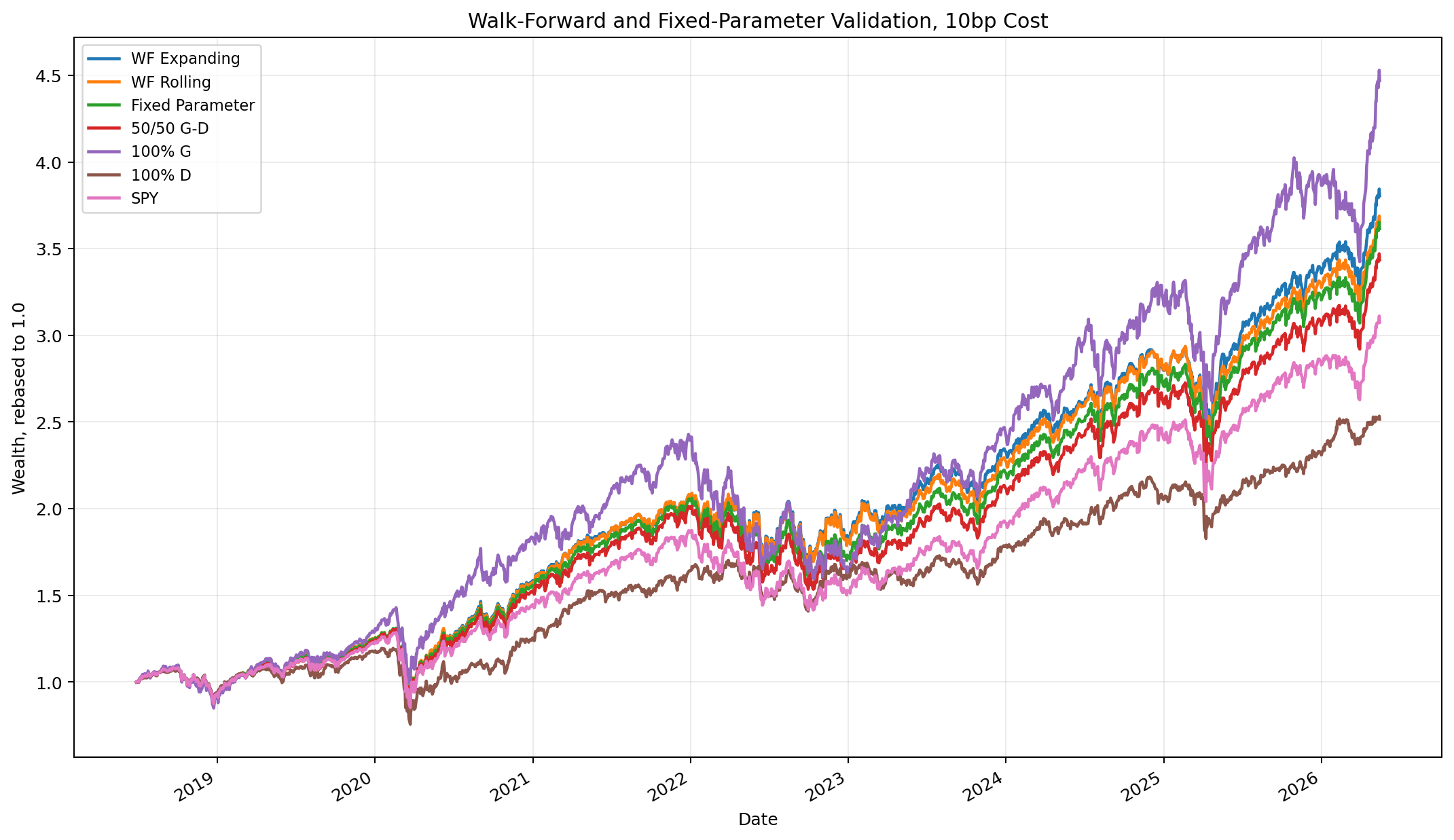}
\caption{OOS Validation Equity Curves, 2018-06-28 to 2026-05-15}
\label{fig:oos}
\end{figure}

\section{Post-2022 Validation}

A post-2022 validation is added because the 2020 COVID crash had an unusual feature: the defensive basket $D$ experienced a larger-than-expected drawdown. To reduce direct dependence on that special event, the OOS validation is repeated from January 3, 2022 to May 15, 2026.

\begin{table}[H]
\centering
\scriptsize
\caption{Post-2022 OOS Validation, 2022-01-03 to 2026-05-15, 10bp Cost}
\label{tab:post2022}
\begin{tabular}{lrrrrrrrr}
\toprule
Method & Final Wealth & CAGR & Vol & Sharpe & Sortino & Max DD & Turnover & Avg $G$\\
\midrule
Smooth Score WF Expanding & 1.86 & \pct{15.30} & \pct{17.53} & 0.90 & 1.22 & \pct{-19.89} & \pct{406.50} & \pct{41.52}\\
Smooth Score WF Rolling & 1.76 & \pct{13.92} & \pct{17.78} & 0.82 & 1.08 & \pct{-21.28} & \pct{390.15} & \pct{45.27}\\
Fixed Parameter & 1.77 & \pct{13.99} & \pct{17.85} & 0.82 & 1.09 & \pct{-22.46} & \pct{83.97} & \pct{47.31}\\
50/50 $G/D$ & 1.72 & \pct{13.23} & \pct{18.02} & 0.78 & 1.02 & \pct{-23.78} & \pct{0.00} & \pct{50.00}\\
100\% $G$ & 1.87 & \pct{15.45} & \pct{23.81} & 0.72 & 0.93 & \pct{-33.92} & \pct{0.00} & \pct{100.00}\\
100\% $D$ & 1.53 & \pct{10.32} & \pct{14.71} & 0.74 & 0.99 & \pct{-17.32} & \pct{0.00} & \pct{0.00}\\
SPY & 1.65 & \pct{12.21} & \pct{17.68} & 0.74 & 0.96 & \pct{-24.50} & \pct{0.00} & --\\
\bottomrule
\end{tabular}
\end{table}

The post-2022 expanding validation is a key piece of evidence for the smooth-score framework. It nearly matches the 100\% $G$ CAGR, \pct{15.30} versus \pct{15.45}, while reducing maximum drawdown from \pct{-33.92} to \pct{-19.89}. This result does not imply that the policy always exceeds 100\% $G$, but it shows that dynamic allocation can deliver growth-like returns with much lower drawdown in the post-2022 window.

\begin{figure}[H]
\centering
\includegraphics[width=0.95\linewidth]{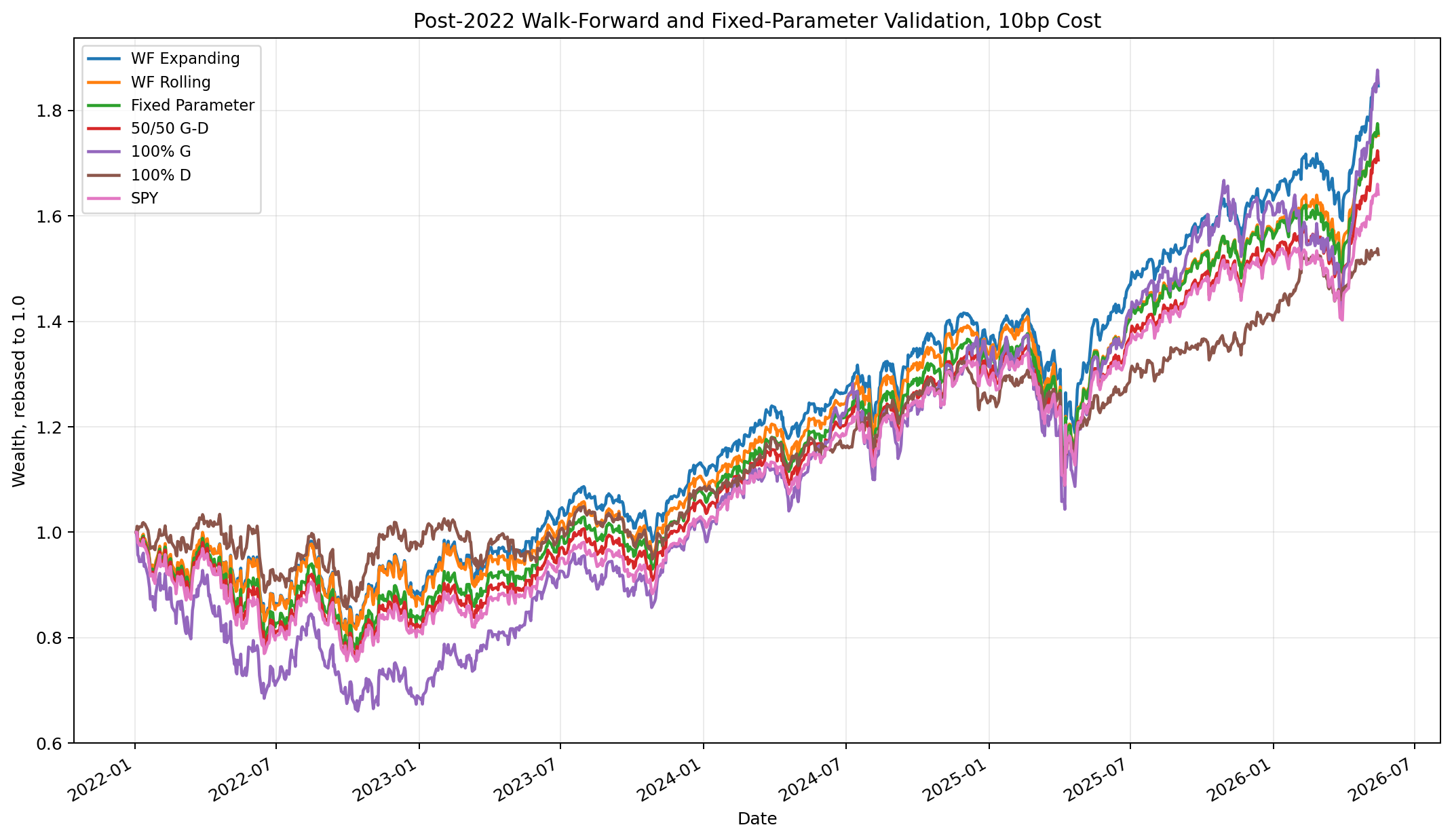}
\caption{Post-2022 OOS Validation Equity Curves, 2022-01-03 to 2026-05-15}
\label{fig:post2022}
\end{figure}

\section{Bond/Credit Incremental Extension}

The preceding sections show that the original smooth score improves risk-adjusted allocation but remains a style-timing policy rather than a new alpha factor. A natural extension is to ask whether bond and credit-market information adds value beyond the already selected score. This section tests that question directly.

The diagnostic branch adds daily credit spread information from FRED and credit ETF proxies. Because HY and IG OAS files currently cover only the recent period beginning in 2023, they are retained for coverage documentation but excluded from the full-window policy test. The deployable full-window bond/credit policy uses BAA/10Y spread information. The direction-normalized credit variables are
\begin{align}
ce_t &= -z(\Delta BAA10Y_{21,t}),\\
cs_t &= z(BAA10Y_t),\\
z(r_t cs_t) &= z(r_t\cdot cs_t),
\end{align}
where $ce_t$ measures credit relief and $cs_t$ measures credit stress.

Two tests are reported. The first is a replacement-style bond/credit score, which uses SPY drawdown depth, credit relief, the $G-D$ 126-day trailing signal, and the rate-relief-by-credit-stress interaction. The second is the stricter incremental test. It keeps the original Best Local structure fixed at
\begin{equation}
\alpha=0.50,\quad \lambda_s=0.50,\quad \lambda_c=0.05,\quad MaxTilt=50\%,\quad \tau_w=0.75,\quad \eta=0.05,
\end{equation}
and adds only
\begin{equation}
\lambda_{credit}ce_t+\lambda_{r\times cs}z(r_tcs_t).
\end{equation}
This second test asks whether bond/credit information improves the original selected score rather than replacing it.

\begin{table}[H]
\centering
\scriptsize
\caption{Bond/Credit Incremental Extension: Main Aligned Comparison, 2017-06-28 to 2026-05-15}
\label{tab:bond-credit-main}
\resizebox{\textwidth}{!}{
\begin{tabular}{lrrrrrrrr}
\toprule
Method & Final Wealth & CAGR & Vol & Sharpe & Max DD & Calmar & Turnover & Avg $G$\\
\midrule
Bond/Credit Smooth Score Best & 4.25 & \pct{17.73} & \pct{19.49} & 0.94 & \pct{-34.50} & 0.51 & \pct{170.28} & \pct{49.62}\\
Bond/Credit Core Only & 4.27 & \pct{17.79} & \pct{19.49} & 0.94 & \pct{-34.48} & 0.52 & \pct{160.35} & \pct{49.55}\\
Old Best + Bond/Credit Incremental & 4.96 & \pct{19.80} & \pct{19.14} & 1.04 & \pct{-31.92} & 0.62 & \pct{410.23} & \pct{43.72}\\
Existing Smooth Score Best Local & 4.76 & \pct{19.24} & \pct{19.29} & 1.01 & \pct{-31.63} & 0.61 & \pct{469.67} & \pct{45.06}\\
50/50 $G/D$ & 4.06 & \pct{17.12} & \pct{19.34} & 0.91 & \pct{-33.59} & 0.51 & \pct{0.00} & \pct{50.00}\\
100\% $G$ & 5.55 & \pct{21.34} & \pct{23.53} & 0.94 & \pct{-34.35} & 0.62 & \pct{0.00} & \pct{100.00}\\
SPY & 3.52 & \pct{15.25} & \pct{18.74} & 0.85 & \pct{-33.72} & 0.45 & \pct{0.00} & --\\
\bottomrule
\end{tabular}}
\end{table}

The replacement-style bond/credit score does not improve on the original smooth-score policy. Its Sharpe ratio is $0.94$, and it underperforms the original Best Local policy in both CAGR and maximum drawdown. This is informative: bond/credit variables are not a superior substitute for the original rate, drawdown, VIX-relief, and crowding score.

The strict incremental branch gives a different result. The selected overlay is
\begin{equation}
\lambda_{credit}=0.10,\quad \lambda_{r\times cs}=0.50.
\end{equation}
It improves the old Best Local policy by \pct{0.56} annualized CAGR and by $0.03$ Sharpe, while reducing turnover by roughly \pct{59.44}. The maximum drawdown is slightly deeper, \pct{-31.92} versus \pct{-31.63}, so the improvement is not a pure drawdown-reduction result. Instead, the credit overlay appears to improve the efficiency of the old signal by reducing average $G$ exposure and lowering turnover while preserving upside participation.

\begin{figure}[H]
\centering
\includegraphics[width=0.95\linewidth]{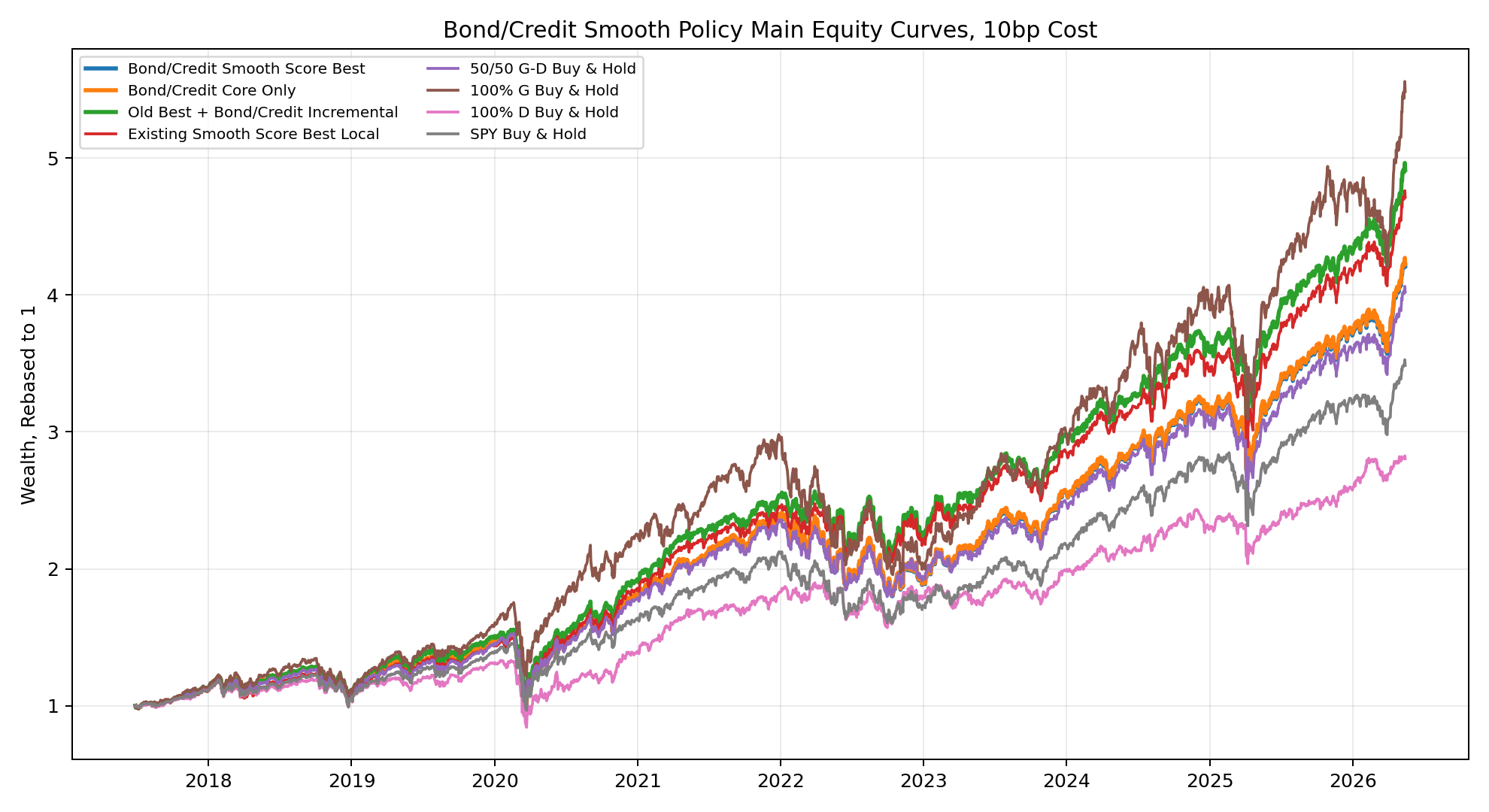}
\caption{Bond/Credit Extension Main Equity Curves, 2017-06-28 to 2026-05-15}
\label{fig:bond-credit-main}
\end{figure}

\begin{table}[H]
\centering
\scriptsize
\caption{Bond/Credit Extension: OOS Validation, 2018-06-28 to 2026-05-15}
\label{tab:bond-credit-oos}
\resizebox{\textwidth}{!}{
\begin{tabular}{lrrrrrrr}
\toprule
Method & Final Wealth & CAGR & Vol & Sharpe & Max DD & Turnover & Avg $G$\\
\midrule
Bond/Credit WF Expanding & 3.44 & \pct{17.02} & \pct{20.14} & 0.88 & \pct{-35.55} & \pct{221.17} & \pct{48.65}\\
Bond/Credit WF Rolling & 3.47 & \pct{17.16} & \pct{20.02} & 0.89 & \pct{-35.55} & \pct{228.98} & \pct{47.12}\\
Bond/Credit Fixed Parameter & 3.47 & \pct{17.15} & \pct{20.42} & 0.88 & \pct{-34.99} & \pct{468.63} & \pct{49.50}\\
Old+Credit WF Expanding & 4.15 & \pct{19.83} & \pct{19.74} & 1.02 & \pct{-32.36} & \pct{388.44} & \pct{43.26}\\
Old+Credit WF Rolling & 4.26 & \pct{20.24} & \pct{19.76} & 1.03 & \pct{-32.36} & \pct{403.31} & \pct{43.11}\\
Old+Credit Fixed Parameter & 4.07 & \pct{19.56} & \pct{19.52} & 1.01 & \pct{-32.54} & \pct{412.32} & \pct{42.74}\\
Existing Smooth Score Best Local & 4.17 & \pct{19.93} & \pct{19.92} & 1.01 & \pct{-31.63} & \pct{449.41} & \pct{44.35}\\
50/50 $G/D$ & 3.46 & \pct{17.10} & \pct{20.00} & 0.89 & \pct{-33.59} & \pct{0.00} & \pct{50.00}\\
\bottomrule
\end{tabular}}
\end{table}

The OOS evidence is consistent with the main-sample evidence. The replacement-style bond/credit score remains weak, while the strict incremental branch is competitive with the existing Best Local policy. Rolling Old+Credit is the strongest of the credit-augmented OOS variants, with \pct{20.24} CAGR and a $1.03$ Sharpe ratio. However, it still does not dominate the old Best Local policy on every metric because its maximum drawdown is slightly worse.

\begin{figure}[H]
\centering
\includegraphics[width=0.95\linewidth]{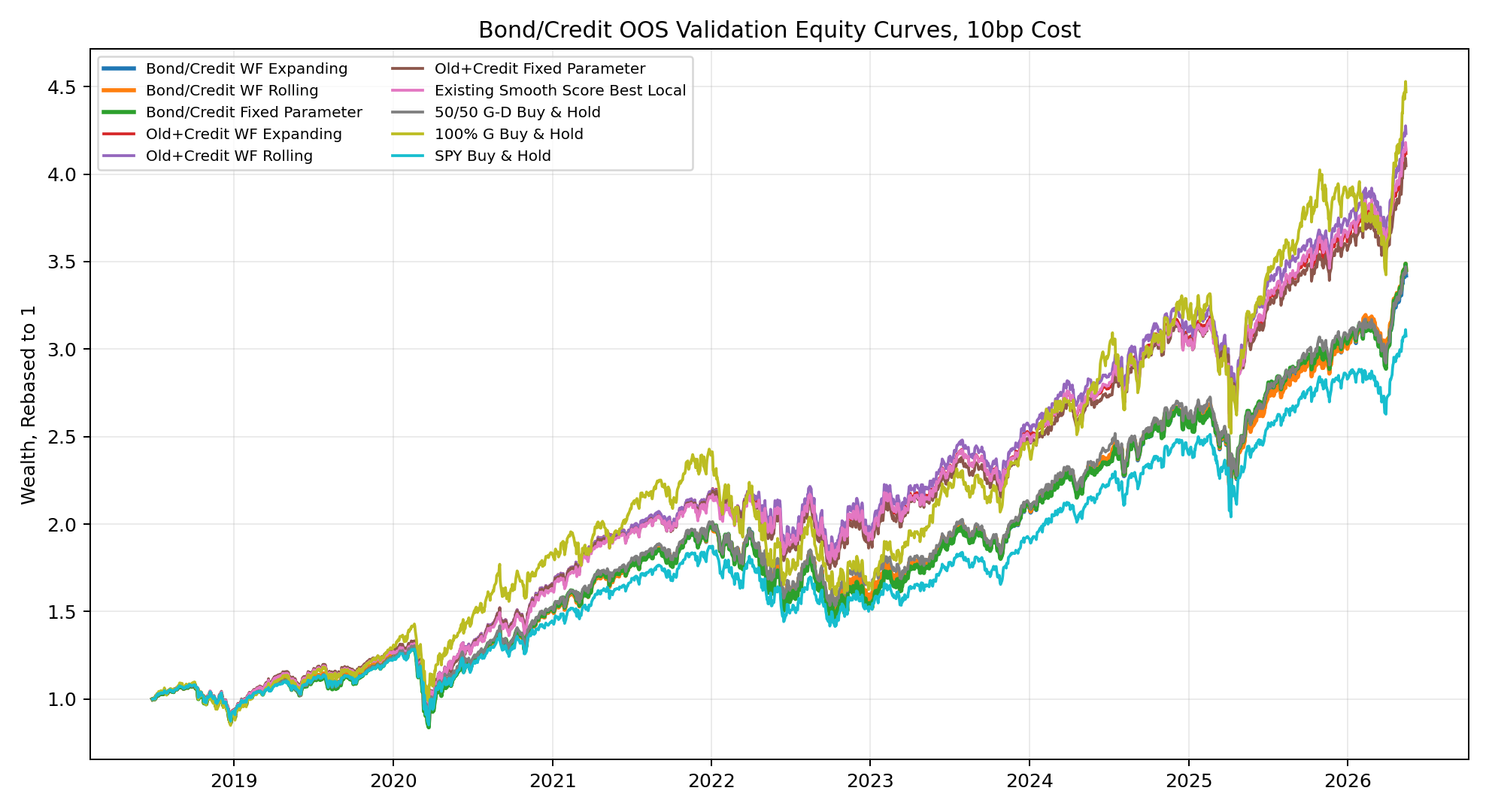}
\caption{Bond/Credit Extension OOS Equity Curves, 2018-06-28 to 2026-05-15}
\label{fig:bond-credit-oos}
\end{figure}

\begin{table}[H]
\centering
\scriptsize
\caption{Bond/Credit Extension: Post-2022 Validation, 2022-01-03 to 2026-05-15}
\label{tab:bond-credit-post2022}
\resizebox{\textwidth}{!}{
\begin{tabular}{lrrrrrrr}
\toprule
Method & Final Wealth & CAGR & Vol & Sharpe & Max DD & Turnover & Avg $G$\\
\midrule
Bond/Credit WF Expanding & 1.73 & \pct{13.39} & \pct{18.33} & 0.78 & \pct{-25.14} & \pct{101.55} & \pct{49.38}\\
Bond/Credit WF Rolling & 1.76 & \pct{13.83} & \pct{17.97} & 0.81 & \pct{-24.55} & \pct{119.91} & \pct{45.97}\\
Bond/Credit Fixed Parameter & 1.76 & \pct{13.95} & \pct{18.42} & 0.80 & \pct{-25.13} & \pct{129.02} & \pct{49.68}\\
Old+Credit WF Expanding & 1.90 & \pct{15.95} & \pct{17.27} & 0.94 & \pct{-19.76} & \pct{365.74} & \pct{38.27}\\
Old+Credit WF Rolling & 1.92 & \pct{16.12} & \pct{17.28} & 0.95 & \pct{-19.45} & \pct{382.49} & \pct{37.51}\\
Old+Credit Fixed Parameter & 1.92 & \pct{16.20} & \pct{17.24} & 0.96 & \pct{-19.68} & \pct{368.67} & \pct{37.65}\\
Existing Smooth Score Best Local & 1.93 & \pct{16.27} & \pct{17.66} & 0.94 & \pct{-19.94} & \pct{455.76} & \pct{40.68}\\
50/50 $G/D$ & 1.72 & \pct{13.23} & \pct{18.02} & 0.78 & \pct{-23.78} & \pct{0.00} & \pct{50.00}\\
100\% $G$ & 1.87 & \pct{15.45} & \pct{23.81} & 0.72 & \pct{-33.92} & \pct{0.00} & \pct{100.00}\\
\bottomrule
\end{tabular}}
\end{table}

The post-2022 validation is particularly useful because it reduces dependence on the COVID crash. In this window, the Old+Credit fixed-parameter variant delivers \pct{16.20} CAGR and a $0.96$ Sharpe ratio, close to the existing Best Local CAGR of \pct{16.27} but with slightly lower volatility and lower turnover. The result supports a narrow conclusion: credit information is useful as an overlay to the original score, especially as a way to moderate exposure and turnover, but it does not justify replacing the original score architecture.

\begin{figure}[H]
\centering
\includegraphics[width=0.95\linewidth]{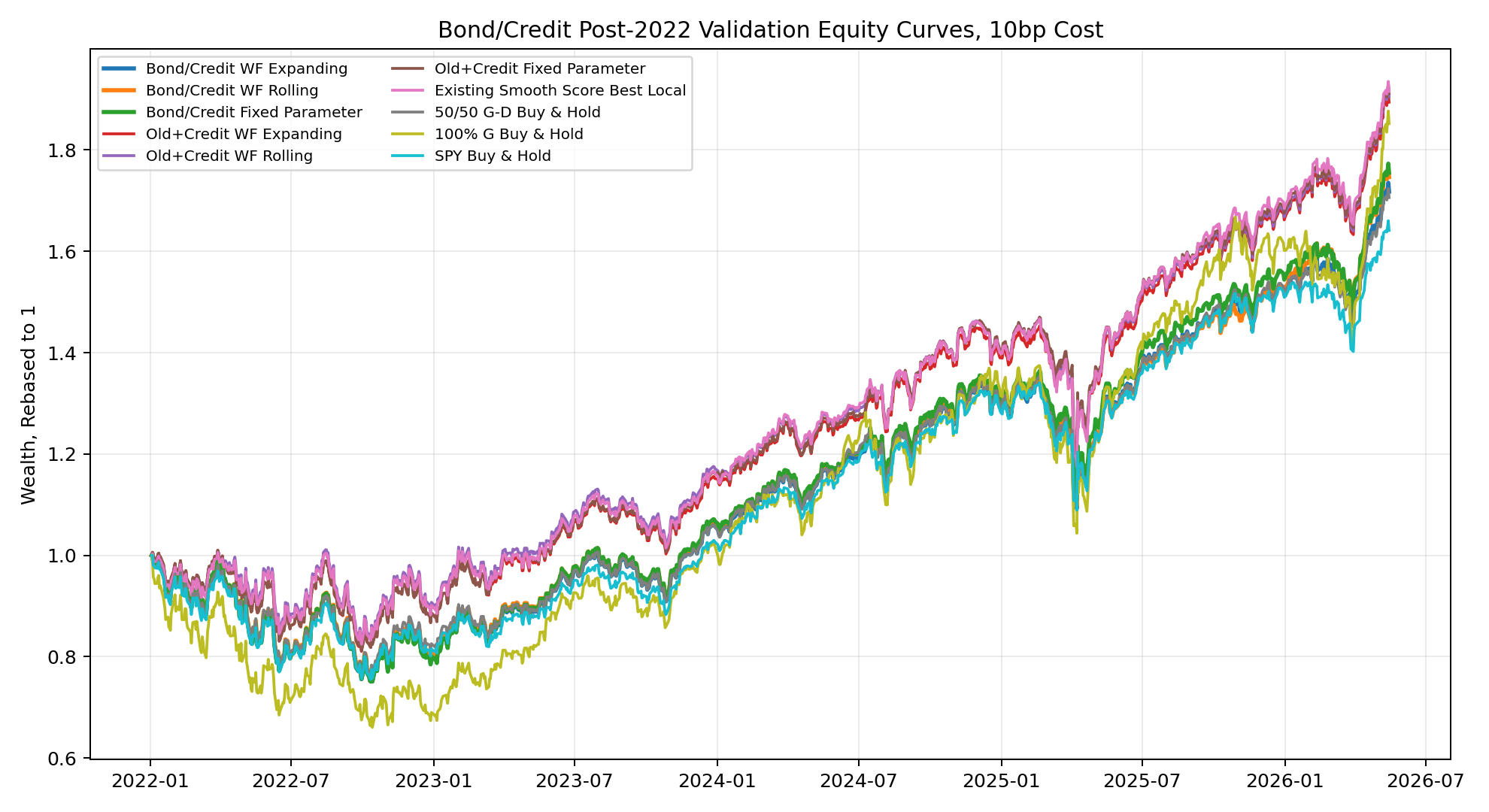}
\caption{Bond/Credit Extension Post-2022 Equity Curves, 2022-01-03 to 2026-05-15}
\label{fig:bond-credit-post2022}
\end{figure}

\section{Score Diagnostics and Yearly Interpretation}

The score-sorting diagnostic checks whether higher scores are associated with higher future $G-D$ returns. It is not used as a trading rule.

\begin{table}[H]
\centering
\caption{Score Sorting Diagnostic}
\label{tab:score-diagnostic}
\begin{tabular}{lrrrrrr}
\toprule
Method & Q1 & Q2 & Q3 & Q4 & Q5 & Q5--Q1\\
\midrule
Selected Smooth Score & \pct{-1.01} & \pct{0.17} & \pct{1.00} & \pct{3.32} & \pct{5.48} & \pct{6.48}\\
Matched Core-only & \pct{0.02} & \pct{-0.25} & \pct{1.16} & \pct{2.48} & \pct{5.57} & \pct{5.56}\\
Fixed-Structure 50\% Tilt & \pct{-0.35} & \pct{-0.04} & \pct{0.51} & \pct{3.63} & \pct{5.22} & \pct{5.57}\\
Matched TNX-only & \pct{0.04} & \pct{0.66} & \pct{1.90} & \pct{2.52} & \pct{3.86} & \pct{3.82}\\
\bottomrule
\end{tabular}
\end{table}

The selected smooth score has the largest Q5--Q1 spread, \pct{6.48}, suggesting stronger ranking ability than TNX-only. The yearly breakdown also clarifies the tradeoff. In 2022, 100\% $G$ loses \pct{-30.61}, while the selected policy loses \pct{-9.53}. In 2023, 100\% $G$ rebounds by \pct{48.31}, while the selected policy returns \pct{29.38}. The policy reduces growth crash exposure, but it gives up some upside in strong growth rallies.

\section{Conclusion}

This paper studies dynamic allocation between growth/technology and defensive income ETF baskets. The factor-attribution evidence shows that $G-D$ is a recognizable style portfolio: positive market beta, strongly negative HML exposure, positive momentum exposure, and no statistically significant alpha after FF5+MOM adjustment.

The smooth-score evidence suggests that a continuous, interpretable score based on rate relief, SPY drawdown, VIX stress relief, and growth-crowding penalties can improve risk-adjusted allocation relative to several relevant benchmarks. The selected 50\% tilt policy improves over 50/50, matched TNX-only, matched core-only, SPY, and vol-matched 100\% $G$. Walk-forward and post-2022 validations provide additional support, especially for drawdown reduction.

The bond/credit extension refines this conclusion. A replacement-style bond/credit score does not improve on the original score, but a strict incremental overlay to the already selected Best Local structure improves CAGR, Sharpe, and turnover in the main aligned window. In OOS and post-2022 validations, the credit overlay remains competitive and occasionally stronger than the original score, though not uniformly dominant. The most defensible interpretation is that credit information has incremental overlay value, not that it should replace the core rate/drawdown/VIX/crowding score.

At the same time, the method does not dominate high static growth exposure in raw CAGR, and turnover remains economically important. The balanced interpretation is therefore that continuous market-state variables can help manage known style exposures, but the evidence does not establish a universal alpha or a production-ready trading system.

\appendix

\section{Bond/Credit Diagnostic Gate}

This appendix summarizes the intermediate diagnostic evidence behind the bond/credit extension. These results are not the final trading conclusion; they are used to decide which credit variables are worth carrying into the deployable smooth-score policy.

\begin{table}[H]
\centering
\scriptsize
\caption{Bond/Credit Main-Effect Gate}
\label{tab:appendix-bc-main-gate}
\resizebox{\textwidth}{!}{
\begin{tabular}{lrrrrr}
\toprule
Variable & Coef 63d & HAC $t$ 63d & Nonoverlap Coef 63d & Direction & Pass\\
\midrule
$d$: SPY drawdown depth & \pct{12.67} & 3.51 & \pct{10.12} & Positive & Yes\\
$g126$: $G-D$ trailing 126d & \pct{-2.17} & -2.25 & \pct{-3.49} & Negative & Yes\\
$ce$: credit relief & \pct{1.31} & 1.70 & \pct{0.14} & Positive & Yes\\
\bottomrule
\end{tabular}}
\end{table}

The main-effect gate shows that SPY drawdown depth remains the strongest continuous signal. The $g126$ variable is negative, consistent with the idea that extended prior growth outperformance can reduce subsequent $G-D$ payoff. Credit relief is weaker than drawdown depth but has the expected positive sign.

\begin{table}[H]
\centering
\scriptsize
\caption{Bond/Credit Interaction Gate}
\label{tab:appendix-bc-interaction-gate}
\resizebox{\textwidth}{!}{
\begin{tabular}{lrrrr}
\toprule
Interaction & Raw Coef 63d & Raw HAC $t$ & TNX-Residual Coef 63d & TNX-Residual HAC $t$\\
\midrule
$r\times cs$: rate relief $\times$ credit stress & \pct{1.71} & 1.51 & \pct{2.07} & 2.36\\
\bottomrule
\end{tabular}}
\end{table}

The interaction gate identifies $r\times cs$ as the key credit interaction. Its residual test is more important than the raw coefficient because it asks whether the interaction explains future $G-D$ after the rate-only component is already absorbed. The residual HAC $t$ of $2.36$ is the reason this term is carried into the strict incremental policy test.

\section{Bond/Credit Policy Grid and Cost Sensitivity}

The bond/credit smooth-score branch tests 793 configurations after adding the strict incremental Old+Credit grid. The selected replacement-style bond/credit score is not the main contribution because it underperforms the existing Best Local score. The strict incremental branch is the relevant incremental evidence.

\begin{table}[H]
\centering
\scriptsize
\caption{Strict Incremental Old+Credit Cost Sensitivity}
\label{tab:appendix-old-credit-cost}
\begin{tabular}{rrrrrrrr}
\toprule
Cost & CAGR & Vol & Sharpe & Max DD & Calmar & Turnover & Final Wealth\\
\midrule
0bp & \pct{20.29} & \pct{19.14} & 1.06 & \pct{-31.87} & 0.64 & \pct{410.23} & 5.14\\
5bp & \pct{20.05} & \pct{19.14} & 1.05 & \pct{-31.89} & 0.63 & \pct{410.23} & 5.05\\
10bp & \pct{19.80} & \pct{19.14} & 1.04 & \pct{-31.92} & 0.62 & \pct{410.23} & 4.96\\
20bp & \pct{19.31} & \pct{19.14} & 1.02 & \pct{-31.97} & 0.60 & \pct{410.23} & 4.78\\
\bottomrule
\end{tabular}
\end{table}

The incremental overlay remains above the old Best Local CAGR under 10bp and remains economically strong under 20bp stress cost. However, transaction cost sensitivity is still material because the strategy's annual turnover remains above \pct{400}.

\bibliographystyle{apalike}
\bibliography{references}

@article{fama1993common,
  title={Common risk factors in the returns on stocks and bonds},
  author={Fama, Eugene F. and French, Kenneth R.},
  journal={Journal of Financial Economics},
  volume={33},
  number={1},
  pages={3--56},
  year={1993}
}

@article{fama2015five,
  title={A five-factor asset pricing model},
  author={Fama, Eugene F. and French, Kenneth R.},
  journal={Journal of Financial Economics},
  volume={116},
  number={1},
  pages={1--22},
  year={2015}
}

@article{carhart1997persistence,
  title={On persistence in mutual fund performance},
  author={Carhart, Mark M.},
  journal={Journal of Finance},
  volume={52},
  number={1},
  pages={57--82},
  year={1997}
}

@article{goyal2008comprehensive,
  title={A comprehensive look at the empirical performance of equity premium prediction},
  author={Goyal, Amit and Welch, Ivo},
  journal={Review of Financial Studies},
  volume={21},
  number={4},
  pages={1455--1508},
  year={2008}
}

@article{campbell2008predicting,
  title={Predicting excess stock returns out of sample: Can anything beat the historical average?},
  author={Campbell, John Y. and Thompson, Samuel B.},
  journal={Review of Financial Studies},
  volume={21},
  number={4},
  pages={1509--1531},
  year={2008}
}

@article{hamilton1989new,
  title={A new approach to the economic analysis of nonstationary time series and the business cycle},
  author={Hamilton, James D.},
  journal={Econometrica},
  volume={57},
  number={2},
  pages={357--384},
  year={1989}
}

@article{guidolin2007asset,
  title={Asset allocation under multivariate regime switching},
  author={Guidolin, Massimo and Timmermann, Allan},
  journal={Journal of Economic Dynamics and Control},
  volume={31},
  number={11},
  pages={3503--3544},
  year={2007}
}

@article{brandt2009parametric,
  title={Parametric portfolio policies: Exploiting characteristics in the cross-section of equity returns},
  author={Brandt, Michael W. and Santa-Clara, Pedro and Valkanov, Rossen},
  journal={Review of Financial Studies},
  volume={22},
  number={9},
  pages={3411--3447},
  year={2009}
}

@article{white2000reality,
  title={A reality check for data snooping},
  author={White, Halbert},
  journal={Econometrica},
  volume={68},
  number={5},
  pages={1097--1126},
  year={2000}
}

@article{hansen2005superior,
  title={A test for superior predictive ability},
  author={Hansen, Peter R.},
  journal={Journal of Business \& Economic Statistics},
  volume={23},
  number={4},
  pages={365--380},
  year={2005}
}

@article{bailey2014probability,
  title={The probability of backtest overfitting},
  author={Bailey, David H. and Borwein, Jonathan M. and L{\'o}pez de Prado, Marcos and Zhu, Qiji Jim},
  journal={Journal of Computational Finance},
  volume={20},
  number={4},
  pages={39--69},
  year={2017}
}

\end{document}